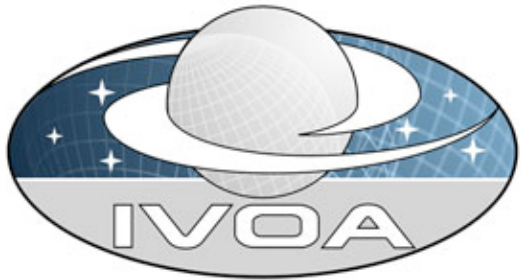

# Simple Cone Search
# Version 1.03

## IVOA Recommendation 22 February 2008

**This version:**
    http://www.ivoa.net/Documents/REC/DAL/ConeSearch-20080222.html
**Latest version:**
    http://www.ivoa.net/Documents/latest/ConeSearch.html
**Previous versions:**
    http://www.ivoa.net/Documents/PR/DAL/ConeSearch-20070914.html
    http://www.ivoa.net/Documents/PR/DAL/ConeSearch-20070628.html
    http://www.ivoa.net/Documents/PR/DAL/ConeSearch-20060908.html

**Working Group:**
    http://www.ivoa.net/twiki/bin/view/IVOA/IvoaDAL
**Editor:**
    Raymond Plante
**Authors:**
    Roy Williams
    Robert Hanisch
    Alex Szalay
    Raymond Plante

---

## Abstract

This specification defines a simple query protocol for retrieving records from a catalog of astronomical sources. The query describes sky position and an angular distance, defining a cone on the sky. The response returns a list of astronomical sources from the catalog whose positions lie within the cone, formatted as a VOTable. This version of the specification is essentially a transcription of the original Cone Search specification in order to move it into the IVOA standardization process.

## Status of this document

This document has been produced by the IVOA Resource Registry Working Group. It has been reviewed by IVOA Members and other interested parties, and has been endorsed by the IVOA Executive Committee as an IVOA Recommendation as of 2007 September 27. It is a stable document and may be used as reference material or cited as a normative reference from another document. IVOA's role in making the Recommendation is to draw attention to the specification and to promote its widespread deployment. This enhances the functionality and interoperability inside the Astronomical Community.

This document was originally published as a document of the US National Virtual Observatory (NVO), available at http://us-vo.org/pubs/files/conesearch.html. This IVOA version is essentially a transcription of that document into the IVOA standards document format. Minor changes have been made to sharpen the specification description where needed and to otherwise fit it into the IVOA standards context; these changes are enumerated in Appendix C. It is the intention of this document that **all existing services compliant with the original NVO document will be compliant with this specification**. Similarly, all future service implementations that are based on this specification are expected to be compliant with the spirit of the original NVO document and practices in the VO current at the time of this writing; thus, existing client applications should work with the new implementations without change.

Comments on this document can be sent to the Data Access Layer Working Group via dal@ivoa.net.

A list of current IVOA Recommendations and other technical documents can be found at http://www.ivoa.net/Documents/.

## Acknowledgments

Cone Search represents the first and arguably the most successful "standard protocol" developed within the Virtual Observatory movement. The editor, therefore, gratefully acknowledges the work of the original authors of the Cone Search specification as well as the numerous data providers who have implemented and continue to support this protocol.

This document has been developed with support from the National Science Foundation's Information Technology Research Program under Cooperative Agreement AST0122449 with The Johns Hopkins University.

## Conformance-related definitions

The words "MUST", "SHALL", "SHOULD", "MAY", "RECOMMENDED", and "OPTIONAL" (in upper or lower case) used in this document are to be interpreted as described in IETF standard, RFC 2119 [RFC 2119].

The **Virtual Observatory (VO)** is general term for a collection of federated resources that can be used to conduct astronomical research, education, and outreach. The **International Virtual Observatory Alliance (IVOA)** is a global collaboration of separately funded projects to develop standards and infrastructure that enable VO applications.

A **Cone** is a circular region on the sky defined by a sky position and a radius around that position. A **Cone Search** is a query for information related to a Cone. The **Cone Search Protocol** refers to the protocol described in this document. A **Cone Search Service** is a service implementation that complies with this standard.

# Contents

---

# 1. Introduction

This specification describes how a provider of an astronomical source catalog can publish that catalog to the Virtual Observatory in such a way that a simple cone search can be done. The data remains in the control of the data provider, served through a web server to the world, but the query profile and response profile are carefully controlled, as described below. It is intended that setting up this web service be simple enough that data providers will not have to spend too much time on it (their funding to support such services is typically small). At the same time, the service implementation and the data it provides can serve as a basis for more sophisticated tools.

This specification does not specify how Cone Search services are implemented, or how the data are stored or manipulated. The concern of this specification is how the data are exposed to the world through well-defined requests and responses.

This specification assumes that the data provider has already selected a catalog of astronomical sources. This catalog can be presented as a single table; it is expected that the table contains several columns.

# 2. Service Interface Requirements

A service implementation that is compliant with this standard must meet the following requirements:

1. the service must respond to a HTTP GET request represented by a URL having two parts:

   ○ A base URL of the form:

   **http://**<server-address>**/**<path>**?[**<extra-GET-arg>**&[...]]**

where *<server-address>* and *<path>* are URI-compliant components indicating the domain address and local location path where the service is deployed. The *<server-address>* may end with one or more locally supported URL arguments, *<extra-GET-arg>*; these arguments are not recognized parts of the Cone Search protocol and thus are treated opaquely by clients of the service as part of the base URL.

Note that when it contains extra GET arguments, the base URL ends in an ampersand, **&**; if there are no extra arguments, then it ends in a question mark, **?**.

> **Examples:**
>     `http://mycone.org/cgi-bin/VOsearch?`
>     `http://adil.ncsa.uiuc.edu/vocone?resolv&issurvey=T&`

Every query to a given cone search service uses the same base URL.

○ Constraints expressed as a list of ampersand-delimited GET arguments of the form:
    *<name>*=*<value>*

where *<name>* is a parameter name specified by this specification and *<value>* is its value. The constraints represent the query parameters which can vary for each successive query. The order of the name-parameter pairs is not significant.

○ The baseURL and constraint list are concatenated to form the query.

○ The set of query constraints must include the following parameters, which are interpreted by the service with the stated meaning:

  ▪ **RA** -- a right-ascension in the ICRS coordinate system for the positon of the center of the cone to search, given in decimal degrees.
  ▪ **DEC** -- a declination in the ICRS coordinate system for the positon of the center of the cone to search, given in decimal degrees.
  ▪ **SR** -- the radius of the cone to search, given in decimal degrees.

> **Example:**
>     `http://mycone.org/cgi-bin/search?RA=180.567&DEC=-30.45&`
>     `SR=0.0125`

○ The query MAY contain the optional parameter, **VERB**, whose value is an integer--either 1, 2, or 3--indicating verbosity which determines how many columns are to be returned in the resulting table. Support for this parameter by a Cone Search service implementation is optional. If the service supports the parameter, then when the value is 1, the response should include the bare minimum of columns that the provider considers useful in describing the returned objects (while still remaining compliant with this standard; see section 2.2 below). When the value is 3, the service should return all of the columns that are available for describing the objects. A value of 2 is intended for requesting a medium number of columns between the minimum and maximum (inclusive) that are considered by the provider to most typically useful to the user. When the VERB parameter is not provided, the server should respond as if VERB=2. If the VERB parameter is not supported by the service, the service should ignore the parameter and should always return the same columns for every request.

○ There may be other parameters in the query, but this document does not specify their meaning or usage. If a query includes an optional parameter, either one specified by this document or not, that is not supported by the service implementation, the service must ignore that parameter.

A query following this syntax represents a request for information on sources located within the specified cone on the sky.

2. The service must respond with an XML document in the VOTable format, that represents a table of astronomical sources whose positions are within the cone (see Appendix A for an example). There may also be other sources outside the cone -- the service implementor may decide on the exact search criterion used internally to the service to select the sources.

The base MIME-type of the HTTP response should be `text/xml`. The more specific form `text/xml;content=x-votable` may optionally be used. The MIME-type, `text/xml;votable` shall also be considered legal (for backward compatibility reasons); however, this value is strongly discouraged as it is not compliant with the MIME-type standard [MIME]. The XML content of the response must be compliant with VOTable schema, version 1.0 [VOTable v1.0] or 1.1 [VOTable v1.1].

> **Editor's Note:**
> The original NVO specification allowed a MIME-type of `text/xml;votable`, thus for backward compatibility this is still allowed but discouraged.

The VOTable MUST comply with these conditions:

○ There must be a single RESOURCE in the VOTable, and that contains a single TABLE.

○ The TABLE must have FIELDS where the following UCD [UCD] values have been set. There must only be one FIELD with each of these UCDs:

■ Exactly one FIELD must have ucd="ID_MAIN", with an array character type (fixed or variable length), representing an ID string for that record of the table. This identifier may not be repeated in the table, and it could be used to retrieve that same record again from that same table.

■ Exactly one FIELD must have ucd="POS_EQ_RA_MAIN", with type double, representing the right-ascension of the source in the ICRS coordinate system.

■ Exactly one FIELD must have ucd="POS_EQ_DEC_MAIN", with type double, representing the declination of the source in the ICRS coordinate system.

○ The VOTable may include an expression of the uncertainty of the positions given in the above mentioned fields to be interpreted either as a positional error of the source positions, or an angular size if the sources are resolved. If this uncertainty is not provided, it should be taken to be zero; otherwise, it may be set for all table entries with a PARAM in the RESOURCE which has a UCD that is set to OBS_ANG-SIZE and has a value which is the angle in decimal degrees. Alternatively, a different value for each row in the table can be given via a FIELD in the table having a UCD set to OBS_ANG-SIZE.

○ There may be other FIELDs in the table. Their specification should include a description, data-type, and UCD.

3. The service must respond with a stubbed version of the VOTable in the case of error. This must happen if the three required parameters (RA, DEC, SR) are not all present, or if their values cannot be parsed to floating-point numbers, or if the numbers are out of range (DEC=91.0, for example). The service may also make an error return if the search radius (give by the SR parameter) is larger than the MaxSR parameter of the Resource Profile (see Section 3).

In the case of error, the service MUST respond with a VOTable that contains a single PARAM element *or* a single INFO element with `name="Error"`, where the corresponding value attribute should contain some explanation of the nature of the error. If an INFO element is used, it must appear as a direct child of one of the following elements:

○ the root VOTABLE element (which is preferred by this document), or
○ the RESOURCE element

If a PARAM element is used, it must appear as a direct child of one of following elements:

○ the RESOURCE element, or
○ a DEFINITION element below the root VOTABLE element (which is discouraged by this document).

> **Editor's Note:**
> It was recognized that this requirement, as it was described in the original NVO specification, is ambiguous about both the element to use for the ERROR message and its location in the document. The VOTable standard allows PARAM and INFO elements to appear in various places within the document. Since several of the different methods have been used in practice by Cone Search service implementations, no attempt is made in this version to correct this ambiguity. All of the above-mentioned locations should be considered legal, and Cone Search service clients should be prepared to look in all of them for an ERROR message. The use of PARAM as a child of DEFINITIONS is discouraged as the DEFINITIONS element was deprecated in VOTable v1.1.

**Example Error Responses:**

Error INFO as child of VOTABLE (preferred)

```
<?xml version="1.0"?>
<!DOCTYPE VOTABLE SYSTEM "http://us-vo.org/xml/VOTable.dtd">
<VOTABLE version="1.0">
   <DESCRIPTION>MAST Simple Cone Search Service</DESCRIPTION>
   <INFO ID="Error" name="Error" value="Error in input RA value: as3f"/>
</VOTABLE>
```

Error PARAM as child of RESOURCE (allowed)

```
<?xml version="1.0"?>
<!DOCTYPE VOTABLE SYSTEM "http://us-vo.org/xml/VOTable.dtd">
<VOTABLE version="1.0">
   <DESCRIPTION>
      HEASARC Browse data service
      Please send inquiries to mailto:request@athena.gsfc.nasa.gov
   </DESCRIPTION>
   <RESOURCE ID="error_resource">
    <PARAM ID="Error" name="Error" datatype="char" arraysize="*"
         value="Invalid data type: text/html"/>
   </RESOURCE>
</VOTABLE>
```

Other conditions may NOT be considered erroneous, for example if a query cone is in the Southern hemisphere and the catalog coverage is in the Northern hemisphere. This type of query is different from an error return; it should return a VOTable as described above, with metadata, but no data records. In particular, a zero value of Search Radius should not return an error condition. This is because an application may be more interested in the metadata than the data, and send a fixed query (for example RA=0&DEC=90&SR=0) simply to discover the fields delivered by the service.

# 3. The Resource Profile

A Cone Search service MUST be described with a Resource Profile that includes the following information. The Profile is composed of named metadata listed below. The format used to encode the profile should be compliant with the publishing standards of the IVOA throughout the time the service is supported by the data provider. The metadata names and values used in the profile encoding should match those given below as closely as possible; where they do not match exactly, they should be consistent with the IVOA metadata conventions in place at any given time and the mapping of names and values actually used and those given below should be well documented.

**Editor's Note:**
    The original NVO specification pre-dates the IVOA standard for
    resource metadata [RM], and so, obviously, some inconsistencies
    with that standard exist. To deal with this, the wording of this

> section has been altered to allow profiles to be constructed according to the latest practices in the IVOA. Appendix B outlines the mapping of the metadata listed below to that in the RM as well as the XML Schema used to render the metadata within a Registry.

Several of the metadata listed below can have values that are hierarchical; this hierarchy should be represented in a manner most appropriate to the format used. When the format does not provide any such mechanism, it is recommended that the value be represented as a strings delimited by dots with the root domain of the value appearing first.

The resource profile consists of the following metadata with the stated definitions:

- **ResponsibleParty**: The data provider's name and email.

- **ServiceName**: The name of the catalog served by the service, for example "IRSA.2MASS.ExtendedSources".

- **Description**: A couple of paragraphs of text that describe the nature of the catalog and its wider context.

- **Instrument**: The instrument that made the observations, for example STScI.HST.WFPC2.

- **Waveband**: The waveband of the observations, with ONE selected from this list: radio, millimeter, infrared, optical, ultraviolet, xray, gammaray.

- **Epoch**: The epoch of the observations, as a free-form string.

- **Coverage**: The coverage on the sky, as a free-form string.

- **MaxSR**: The largest search radius, given in decimal degrees, that will be accepted by the service without returning an error condition. A value of 180.0 indicates that there is no restriction.

- **MaxRecords**: The largest number of records that the service will return.

- **Verbosity**: True or false, depending on whether the service supports the VERB keyword in the request.

- **BaseURL**: The base URL for the service as described in Section 2.

The service will be considered published to the VO if the profile has been added to an IVOA Registry according to the IVOA standards and conventions at the time the service is made available, TOGETHER with maintaining the web service that is described by the profile in compliant order.

## Appendix A: Sample VOTable Response

This is a sample of a legal response to a Cone Search query.

## Example VOTable Response:

```
<?xml version="1.0"?>
<!DOCTYPE VOTABLE SYSTEM "http://us-vo.org/xml/VOTable.dtd">
<VOTABLE version="1.0">
  <DEFINITIONS>
    <COOSYS system="eq_FK5" equinox="2000" />
  </DEFINITIONS>

  <RESOURCE ID="T9001">
    <DESCRIPTION>
      HEASARC Browse data service
      Please send inquiries to mailto:request@athena.gsfc.nasa.gov
    </DESCRIPTION>
    <PARAM ID="default_search_radius" ucd="OBS_ANG-SIZE" datatype="double"
          value="0.0516666666666667" />

    <TABLE ID="heasarc_first_9001">
      <DESCRIPTION> Faint Images of the Radio Sky at Twenty cm Source Catalog (

      <FIELD name="unique_id" datatype="char" arraysize="*"  ucd="ID_MAIN">
        <DESCRIPTION> Integer key </DESCRIPTION>
      </FIELD>

      <FIELD name="name" datatype="char" arraysize="*"  >
        <DESCRIPTION> FIRST Source Designation </DESCRIPTION>
      </FIELD>

      <FIELD name="ra" datatype="double" unit="degree" ucd="POS_EQ_RA_MAIN">
        <DESCRIPTION> Right Ascension </DESCRIPTION>
      </FIELD>

      <FIELD name="dec" datatype="double" unit="degree" ucd="POS_EQ_DEC_MAIN">
        <DESCRIPTION> Declination </DESCRIPTION>
      </FIELD>

      <FIELD name="flux_20_cm" datatype="double" unit="mJy" >
        <DESCRIPTION> Peak 1.4GHz Flux Density (mJy) </DESCRIPTION>
      </FIELD>

      <FIELD name="flux_20_cm_error" datatype="double" unit="mJy" >
        <DESCRIPTION> Estimated rms in at Source (mJy) </DESCRIPTION>
      </FIELD>

      <FIELD name="int_flux_20_cm" datatype="double" unit="mJy" >

        <DESCRIPTION> Integrated 1.4GHz Flux Density (mJy) </DESCRIPTION>
      </FIELD>

      <DATA>
        <TABLEDATA>
<TR>
  <TD>384559</TD><TD>FIRST J120002.6+595708</TD>
  <TD>180.0110042</TD><TD>59.9523889</TD>
  <TD>     1.11</TD><TD> 0.139</TD><TD>     1.14</TD>
</TR>
<TR>
  <TD>385094</TD><TD>FIRST J120025.3+600103</TD>
  <TD>180.1057250</TD><TD>60.0175556</TD>
  <TD>     2.89</TD><TD> 0.142</TD><TD>     2.56</TD>
</TR>
<TR>
```

```
      <TD>384928</TD><TD>FIRST J120018.1+600236</TD>
      <TD>180.0755500</TD><TD>60.0434750</TD>
      <TD>    19.38</TD><TD> 0.145</TD><TD>    19.23</TD>
   </TR>
   <TR>
      <TD>384490</TD><TD>FIRST J115959.4+600403</TD>
      <TD>179.9978875</TD><TD>60.0677083</TD>
      <TD>     1.01</TD><TD> 0.147</TD><TD>     1.20</TD>
   </TR>
          </TABLEDATA>
        </DATA>
      </TABLE>

    </RESOURCE>
</VOTABLE>
```

# Appendix B: Current Practices for Representing Resource Profiles

## B.1. Mapping for Resource Profile Metadata to the RM

As mentioned in an Editor's Note in Section 3, the original NVO specification pre-dated the IVOA standard for resource metadata known as the RM [RM]. This section indicates how the resource profile metadata defined in this specification maps to the metadata defined in the RM.

| Cone Search Metadatum | RM Metadatum |
|---|---|
| **ResponsibleParty** | Publisher, Contact.Email |
| **ServiceName** | Title |
| **Description** | Description |
| **Instrument** | Instrument |
| **Waveband** | Coverage.Spectral |
| **Epoch** | Coverage.Temporal.StartTime, Coverage.Temporal.StopTime[1] |
| **Coverage** | Coverage.Spatial |
| **MaxSR** | Service.MaxSearchRadius |
| **MaxRecords** | Service.MaxReturnRecords |
| **Verbosity** | *n/a*[2] |
| **BaseURL** | Service.BaseURL |

[1]The notion of the epoch the observations is captured in the RM as the temporal coverage. The notion of the equinox of the observational positions is captured part of the RM's Coverage.Spatial.

[2]As this concept is not covered by the RM, it should be considered service-specific capability metadata.

## B.2. VOResource (pre-v1.0) Schema Extension for Cone Search

## Services

Just prior to the adoption of IVOA standard for registry interfaces, v1.0 [RI], resource descriptions were encoded using the VOResource XML Schema, v0.10 and its family of extension schemas. The extensions used to specifically describe Cone Search services were VODataService, v0.5 and ConeSearch, v0.3. See the embedded documentation in each of these schemas for the precise definitions and usage of the metadata encodable through them.

The following table enumerates the mapping of resource profile metadata defined in this specification with those defined in the XML schemas.

| Cone Search Metadatum | VOResource Metadatum | |
|---|---|---|
| | Schema Name | XPath Name |
| **ResponsibleParty** | VOResource | `curation/publisher,`<br>`curation/contact/email` |
| **ServiceName** | VOResource | `title` |
| **Description** | VOResource | `content/description` |
| **Instrument** | VOResource | `instrument` |
| **Waveband** | VODataService | `coverage/spectral/waveband` |
| **Epoch** | VODataService | `coverage/temporal`<br>`/startTime,`<br>`coverage/temporal/stopTime`[1] |
| **Coverage** | VODataService | `coverage/spatial`[2] |
| **MaxSR** | ConeSearch | `capability/maxSR` |
| **MaxRecords** | ConeSearch | `capability/maxRecords` |
| **Verbosity** | ConeSearch | `capability/verbosity` |
| **BaseURL** | VOResource | `interface/accessURL` |

[1] The notion of the epoch the observations is captured in the schema inside coverage/temporal. The notion of the equinox of the observational positions is captured within coverage/spatial/region, such as in coverage/spatial /region[@xsi:type='vs:Circle']/coordFrame.

[2] the coverage/spatial element encodes its information as a complex set of child elements.

## B.3. VOResource (v1.0) Schema Extension for Cone Search Services

With the expected adoption of the IVOA standard for Registry Interface [RI], resources are described using the VOResource schema, v1.0 [VOR] and its family of extensions. Cone Search services are specifically described using the VODataService, v1.0, and ConeSearch, v1.0, extension schemas. Coverage information is encoded using the Space-Time Coordinates (STC) schema [STC].

The following table enumerates the mapping of resource profile metadata defined in this specification with those defined in the XML schemas.

| Cone Search Metadatum | VOResource Metadatum | |
| --- | --- | --- |
| | Schema Name | XPath Name |
| **ResponsibleParty** | VOResource | `curation/publisher,`<br>`curation/contact/email` |
| **ServiceName** | VOResource | `title` |
| **Description** | VOResource | `content/description` |
| **Instrument** | VOResource | `instrument` |
| **Waveband** | VODataService | `coverage/waveband` |
| **Epoch** | VODataService | `coverage/stc:STCResourceProfile`<br>`/stc:AstroCoordArea`[1] |
| **Coverage** | VODataService | `coverage/stc:STCResourceProfile`<br>`/stc:AstroCoordArea`[1] |
| **MaxSR** | ConeSearch | `capability/maxSR` |
| **MaxRecords** | ConeSearch | `capability/maxRecords` |
| **Verbosity** | ConeSearch | `capability/verbosity` |
| **BaseURL** | VOResource | `capability/interface[@role='std']`<br>`/accessURL` |

[1] In STC, coverage on the sky and in time are described in an integrated way within the stc:AstroCoordArea element. The notion of the equinox of the observational positions is captured within stc:AstroCoordSystem element.

Subsequent versions of this document should include the ConeSearch extension schema as a formal part of the specification.

# Appendix C: Change History

## Appendix C.1: Changes From the Original NVO Specification Document

- References to the original HTML document have been replaced with references to this IVOA specification.
- Replaced references to "curator" with "data provider" or similar wording.
- Replaced references to the NVO with references either to the IVOA or this specification, as appropriate.
- Ambiguous language like "perhaps" has been replaced with more definitive wording where appropriate. Use of the word "will" is replaced with "must" and "can" with "may", in accordance with the definitions set in the preface.
- Grammatical and spelling corrections have been made.
- The content is organized into sections typical of an IVOA spec.
- Description of the URL syntax is sharper, borrowing language from the SIA specification [SIA]. This includes:
  - more explicitly specifying the form of the URL
  - sharpening the definition of the input parameters
  - indicating that parameter order is not significant
  - stating explicitly that unsupported optional arguments should be ignored.
  - adding examples
  - re-ordering information for improved flow.

- The version of VOTable supported is explicitly stated.
- Whereas the NVO version describes the parameter with ucd="OBS_ANG-SIZE" as "an expression of the opening angle of the cones", this version describes it specifically as "an expression of the uncertainty of the positions".
- A note has been added to recognize the ambiguity in the location of the ERROR parameter.
- The general description of the resource profile has been altered to allow rendering of the metadata to change according to the standards and conventions of the IVOA.
- An editor's note has been added that makes reference to the RM document [RM].
- A requirement that **MaxSR** be given in decimal degrees has been added.
- For the **BaseURL** resource profile metadatum, the example has been replaced with a reference to the BaseURL syntax description.
- An appendix has been added to describe the current practice for registering Cone Search services.

## Appendix C.2: Changes from Previous IVOA Versions

### Changes from v1.01

- eliminated the choice of encoding an ERROR response in a PARAM that is a direct child of VOTABLE as this is not legal in the VOTable schema.
- allowed the use of the INFO element for error messages.
- In examples, made sure PARAM elements have datatype attributes
- Corrected wording to be definitive that positions are given in the ICRS coordinate system.
- other typos

### Changes from v1.02:

- converted to Recommendation

### Changes from v1.00

- various typos
- clarified description of VERB parameter:
  - clarified what is meant by optional for client and server
  - clarified the meaning of the values
- Explicitly mention the 3 legal locations for ERROR messages
- refer to string types as character arrays, as per the VOTable std.
- prefers text/xml;content=x-votable over text/xml;votable.
- added examples of error message, legal response in appendix

## References

**[RFC 2119]**
    Bradner, S. 1997. *Key words for use in RFCs to Indicate Requirement Levels*, IETF RFC 2119, `http://www.ietf.org/rfc/rfc2119.txt`
**[MIME]**

Last modified: Fri Sep 14 17:03:48 2007